\documentclass[12pt,preprint]{aastex}
\usepackage{epsfig}

\begin{document}

\title{A {\em Chandra} Spectroscopic Survey of Persistent Black 
Hole Candidates} 

\author{Wei Cui\altaffilmark{1}, Y.X. Feng\altaffilmark{1},
S.~N.~Zhang\altaffilmark{2}, M.W. Bautz\altaffilmark{3}, 
G.P. Garmire\altaffilmark{4}, and N.S. Schulz\altaffilmark{3}} 

\altaffiltext{1}{Department of Physics, Purdue University, West 
Lafayette, IN 47907}

\altaffiltext{2}{Physics Department, University of Alabama in 
Huntsville, Huntsville, AL 35899}

\altaffiltext{3}{Center for Space Research, Massachusetts 
Institute of Technology, Cambridge, MA 02139}

\altaffiltext{4}{Department of Astronomy and Astrophysics, The
Pennsylvania State University, University Park, PA 16802}

\begin{abstract}

We present results from observations of persistent black hole
candidates with the High Energy Transmission Gratings aboard
the {\em Chandra X-ray Observatory}. The sources include LMC X-1, 
LMC X-3, GRS 1758-258, and Cyg X-1. Along with the published
results on 1E1740.7-2942, we have completed a high-resolution
spectroscopic survey of such systems. The observed X-ray spectra 
of LMC X-1 and LMC X-3 show no prominent discrete features, 
while absorption edges (Mg K and Si K) are detected in the 
spectrum of GRS 1758-258. The edges are likely to be of 
interstellar origin. In most cases, the X-ray continuum can be 
described well by models that are often adopted in 
low-resolution studies of black hole candidates: a 
multi-temperature disk spectrum plus a Comptonization component. 
However, the relative contribution of the two components varies
greatly among different sources. For instance, only the disk 
component is present for LMC X-1 and GRS 1758-258, while the 
Comptonized component is required for other sources. We
discuss general issues related to obtaining disk parameters 
from modeling X-ray continuum. 

\end{abstract}


\section{Introduction}

Prior to the launch of {\em Chandra X-ray Observatory} (Chandra), 
emission lines and absorption lines or edges had already been 
detected in the X-ray spectrum of a few black hole candidates 
(BHCs), with X-ray spectrometers of low to moderate resolution 
(e.g., Barr et al. 1985; Kitamoto et al. 1990; Done et al. 1992; 
Ebisawa et al. 1996; Cui et al. 1998; Ueda et al. 1998; Cui et 
al. 2000; Feng et al. 2001). The features often appear in the 
energy range 6--8 keV and are, therefore, attributed to emission 
or absorption processes involving iron K-shell electrons. 
Although the exact location of the emitting or absorbing material 
is often debatable, there is evidence that, at least for some 
black hole candidates, the 
observed iron $K\alpha$ line appears to originate in the innermost 
region of the accretion disk, very close to the central black hole. 
If this proves to be the case, the profile of the line would be 
distorted by the strong gravitational field of the hole (Fabian 
et al. 1989; Tanaka et al. 1995) and could be carefully modeled 
to constrain the 
intrinsic properties of the hole, such as its mass and spin (Laor 
1991; Bromley et al. 1997).

Two major advances were brought about by the CCD spectrometer 
aboard {\em ASCA}. First, several pairs of Doppler-shifted 
emission lines were observed and attributed to the relativistic 
jets in SS 433 (Kotani et al. 1996). A recent observation of the 
source with {\em Chandra} confirmed the presence of these lines 
from the jets and provided many more details (Marshall et al. 2002), 
thanks to the improved spectral resolution of the {\em High-Energy 
Transmission Gratings} (HETG) aboard. The studies of such lines 
have proven fruitful in gaining insights into the dynamics and 
physical conditions of the jets in SS 433 (Marshall et al. 2002; 
Kotani et al. 1996). Similar Doppler shifted lines also seem 
to exist in the spectrum of another well-known jet source, 
1E1740.7-2942, based on a recent {\em Chandra} HETG observation
(Cui et al. 2001). Second, the {\em ASCA} observations of 
micro-quasars GRS 1915+105 and GRO J1655-40 revealed the presence 
of absorption lines (Ueda et al. 1998; Kotani et al. 2000). The 
results are confirmed by a recent {\em Chandra} observation of 
GRS 1915+105 (Lee et al. 2001). The absorption lines were 
interpreted as resonant absorption lines due to highly ionized 
(helium or hydrogen like) ions in a non-spherical configuration. 
There was speculation as whether such lines were characteristic 
of microquasars. This is clearly not the case, as shown by recent 
{\em Chandra} observations of Cyg X-1 (Schulz et al. 2002; Marshall 
et al. 2001). The obtained HETG spectra of Cyg X-1 show numerous 
narrow absorption lines due to H-like or He-like ions of various 
elements. Therefore, the phenomenon may be common for BHCs.

The presence of energetic electrons in the vicinity of accretion 
disks has always been a key ingradient in
nearly all attempts to model the X-ray continuum of a BHC. The
observed power-law spectrum at high energies ($\gtrsim 10$ keV)
is invariably attributed to inverse Comptonization of soft photons
by such electrons in an optically thin but geometrically thick
configuration (review by Tanaka \& Lewin 
1995). At low energies, contribution from the accretion
disk becomes important and often dominant. The $\alpha$-disk
solution (Shakura \& Sunyaev 1973) is often adopted to describe the 
observed spectrum of the disk emission; this has led to the 
formulation of the popular multi-color disk (MCD) model (Mitsuda et 
al. 1984). The model has worked satisfactorily in describing 
low-resolution data, but so have other models (such as a simple 
blackbody). The high-resolution data from {\em Chandra} (and 
{\em XMM-Newton}) may now allow us to see subtle differences in
the shape of the continuum as predicted by various models.

The MCD model contains fundamental physical information
about the accretion disk, such as its temperature (distribution)
and the distance from the inner edge of the disk to the black hole.
Existing observational evidence shows that, under certain 
conditions, the accretion disk extends down to the last 
(marginally) stable orbit (Tanaka \& Lewin 1995 and references 
therein). Therefore, modeling emission from the disk may lead to a 
determination of the radius of the last stable orbit and thus the 
properties of the black hole (e.g., its mass and angular momentum; 
Zhang, Cui, \& Chen 1997; Tanaka \& Lewin 1995). However, the local 
X-ray spectrum of an accretion disk is not a perfect blackbody, 
because in the hot inner portion of the disk the opacity is 
primarily due to electron scattering as apposed to free-free 
absorption. The spectrum can instead be described by a ``diluted'' 
blackbody (Ebisuzaki, Hanawa, \& Sugimoto 1984), with its 
temperature higher than the effective temperature. Attempts have 
been made to quantify such effects theoretically in terms of a
spectral hardening factor (Shimura \& Takahara 1995), but the 
results are still quite uncertain. In particular, it has been 
shown that the hardening factor seems to vary with mass accretion 
rate (Merloni, Fabian, \& Ross 2000). On the other hand,
the derivation of disk properties depends critically on this factor.
The question becomes whether we can determine it observationally.

In this paper, we present results from a survey of persistent BHCs 
with the HETG. The motivations for conducting the survey included: 
(1) using emission or absorption lines, if present, to study the 
physical properties of the emitting region and its environment; 
(2) obtaining profiles of those emission lines that might originate 
in the accretion disk, with a goal of inferring the properties of 
the black hole; (3) directly detecting X-ray emission from the jets 
in microquasars; and (4) better characterizing the disk component 
of the continuum, making use of good low-energy sensitivity as well 
as excellent energy resolution of the HETG. The survey might also 
have provided data for systematic studies of line production 
in persistent BHCs. The results are likely to be relevant to 
transient BHCs, which constitutes the majority.

\section{Sources and Observations}

The sample of sources for the survey originally included LMC X-1, 
LMC X-3, Cyg X-1, and 1E1740.7-2942. Cyg X-1 was subsequently 
observed a few more times. The results on spectral lines and 
edges in Cyg X-1 have been presented elsewhere (Schulz et al 2002; 
Marshall et al. 2001). Here, we used a subset of data from the 
first 
observation (see Schulz et al. 2002 for details) to characterize 
the continuum. 1E1740.7-2942 was omitted from this study because 
the results have also been published (Cui et al. 2001). On the 
other hand, we re-analyzed an HETG observation of GRS1758-258 
(with data taken from the public archive) and the results are 
presented here for completeness (see also Heindl \& Smith 2001).

LMC X-1 and LMC X-3 are both located in the Large Magellanic Cloud 
(LMC), about 55 kpc away. Unlike most BHCs, these two sources are 
usually seen to be in the ``soft state'' where the soft X-ray 
($<$ 10 keV) flux is relatively high and the X-ray spectrum is 
relatively soft. Evidence for the presence of an iron K$_{\alpha}$ 
line in the X-ray spectrum exists for LMC X-1 (Ebisawa et al. 1989; 
Schlegel et al. 1994) and LMC X-3 (Nowak et al. 2001), but it is 
not very convincing in either case. In the former, the line
appears to be weak and narrow, while in the latter only a broad 
line is allowed by the observations.

LMC X-1 was observed for about 20 ks on 2000 January 16. We adopted 
the default time exposure (TE) mode for the observation. The zeroth 
order image is severely piled up (i.e., multiple photons hit the 
same event detection cell of the ACIS within one readout frame), 
because of the brightness of LMC X-1. In fact, the pile-up
effect is so severe that the events at the source position were 
rejected by the on-board processing system (and thus not transmitted 
back), leaving a ``hole'' in the image. Because we were mostly 
interested in using higher-order data to construct a high-resolution 
spectrum of the source, we sacrificed the imaging capability for 
better statistics in the dispersed spectra. 
 
LMC X-3 was also observed for about 20 ks on 2000 Febuary 29. In 
this case, we adopted an alternating exposure mode to obtain some 
zeroth-order data that is nearly pile-up free. The observation 
was conducted with eight regular exposure frames (with a readout
time of 3.2 sec) followed by one short exposure frame (with a 
readout time 0.3 sec). Although the use of alternating exposure 
modes reduces the observing efficiency, the loss is almost 
negligible in this case. Unlike LMC X-1, which is a relatively 
steady source, LMC X-3 seems to vary in a quasi-periodic fashion
(with a characteristic timescale of 199 days; Cowley et al. 1991;
see also the light curves\footnote{http://xte.mit.edu/ASM\_lc.html} 
from the All-Sky Monitor (ASM) on {\em Rossi X-ray Timing Explorer} 
(RXTE)). The {\em Chandra}
observation took place when the flux of the source was quite low,
as seen by the ASM, but it was still high enough to cause significant 
pile-up in the zeroth-order image.

GRS 1758-258 is a persistent X-ray source near the center of our
own galaxy. It is considered a black hole candidate only because 
of its similarities to Cyg X-1 both in terms of the observed spectral 
and temporal X-ray properties. The most remarkable property known of 
this source is the presence of persistent, two-sided radio jets,
reminiscent of extragalactic radio-loud quasars (Rodr\'\i guez
et al. 1992). Such a comparison also supports the suggestion that 
GRS 1758-258 is an X-ray binary that contains an accreting black 
hole. The source shows uncanny similarities to 1E1740.7-2942 
(Mirabel \& Rodr\'\i guez 1999), both in terms of the observed 
X-ray properties and radio properties. Perhaps, the two systems 
are indeed intrinsically similar. 

GRS 1758-258 was observed with the HETG for about 30 ks on 2001 
March 24. A $\frac{1}{2}$-subarray mode was adopted to reduce the 
readout time and thus to minimize expected pile-up in the zeroth-order
image. Although the use of the subarray mode raises the lower 
energy threshold of the dispersed spectra, the effects are minimal
here because it is still the interstellar absorption that limits
the detection of low-energy photons. Like 1E1740.7-2942, 
GRS 1758-258 is thought to be relatively steady and seems always  
to be in the ``hard state'' (Tanaka \& Lewin 1995 and references 
therein). When this {\em Chandra} observation was carried out, 
the X-ray flux of the source was unusually low (Smith et al. 2001).

Cyg X-1 is the first system that was dynamically determined to 
contain a black hole. It spends most of the time in the hard state. 
Once every a few years, it undergoes a transition to the soft state, 
and remains there for weeks to months before returning to the hard
state. An iron K$_{\alpha}$ line has been observed in the X-ray 
spectrum of Cyg X-1 (Barr et al. 1985; Done et al. 1992; Ebisawa et 
al. 1996; Cui et al. 1998). The line seems to be relatively broad 
and strong (however, see Ebisawa et al. 1996). In some cases, it is 
thought to be significantly red-shifted (Barr et al. 1985). An 
interesting evolution of the line properties has been observed 
during the spectral state transition, which seems to suggest a 
physical origin of the line in the accretion disk (Cui et al. 1998).

Cyg X-1 was observed for about 11 ks on 1999 October 19. An 
alternating exposure mode was also used here, with five long frames 
followed by one short frame (with a readout time of 0.3 sec). The 
pile-up effects are extremely severe in this case, not only 
affecting the zeroth-order image but also the dispersed spectra. 
More details regarding this observation can be found in Schulz et 
al. (2002), along with the procedures to deal with complications 
caused by pile-up and the results on the detection of discrete 
spectral features. Here, we only use the short-frame data to 
construct a reliable continuum for comparison with other sources.

\section {Analysis and Results}

The HETG consists of two grating assemblies: the medium-energy
grating (MEG; 0.4--5 keV or 2.5--31 \AA) and high-energy 
grating (HEG; 0.9--10 keV or 1.2--14 \AA). The two gratings 
are offset by a small angle to avoid the overlap of their
dispersed spectra. The resolving power reaches as high as about
1000 for both gratings. The separation of various orders is
achieved by utilizing the intrinsic energy resolution of the 
ACIS CCDs.

With the standard software package {\em CIAO 2.2} (provided by 
the {\em Chandra} X-ray center), we first processed the Level 1 
data to produce Level 1.5 and then Level 2 products. The Level 2
data was then reduced and analyzed, again using {\em CIAO}, to
construct images and spectra of the sources. Spectral modeling 
was carried out with {\em XSPEC 11.1.0}.

\subsection{Imaging Analysis}

Fig.~1 shows images of the central region of the field of view 
for all observations. The readout trace is visible (quite bright
in the Cyg X-1 image), cutting through the position of the source. 
For LMC X-3 and Cyg X-1, the images were generated from the 
long-frame data, so a ``valley'' (as opposed to a peak) is seen 
at the position of the source due to severe pile-up. The same is
true for LMC X-1, as expected. In all cases, a diffuse ``halo'' 
is present around the source. The halo is spherically 
symmetric and is centered right on the source. It is almost 
certainly due to the
scattering of X-rays from the source by interstellar dust along
the line-of-sight. Note that the halo emission is also apparent 
in the dispersed spectra for Cyg X-1 because the source is so
bright. Since photons of lower energies are deflected
by larger angles, the size of the halo is determined by two
competing factors: dust scattering and photo-electric absorption;
the larger the column density, the more likely a photon is 
scattered out of the line-of-sight (strengthing the halo) or 
absorbed (weakening the halo). The scattering halos are known 
to exist around many X-ray sources and have proven useful for 
studying the spatial and size distributions of dust grains in 
our Galaxy (e.g., Mauche \& Gorenstein 1986; Predehl \& Schmitt 
1995; Witt, Smith, \& Dwek 2001). {\em Chandra} now makes it 
possible to directly image the 
phenomenon. The observations should, in principle, allow us to 
derive spatially resolved energy distribution of photons in the 
halo and thus to test theoretical models at a more detailed 
level. Such an investigation is beyond the scope of this paper.

A natural product of the observations is a precise determination 
of the position of each source, thanks to the unprecedented 
spatial resolution of {\em Chandra}. For LMC X-3 and Cyg X-1,
the zeroth-order image from the short-frame data is nearly 
free of pile-up effects. It was used to determine the peak 
positions of linear count profiles across the directions of 
right accension (RA) and declination (Dec), which directly 
yields the source coordinates. However, the short-frame data
suffers from poor statistics. The uncertainties on the positions 
are, therefore, significantly larger than those derived from 
the long-frame data (which is described below).

Although the zeroth-order image from the long-frame data (or
regular frames for LMC X-1 and GRS 1758-258) may be severely 
piled up, the position of a source can be determined from the 
intersection between the readout trace and dispersed MEG or HEG
image. Since the readout trace is caused by the process in
which photons from the source hits the CCD chip during the 
frame transfer operation, the source must be located somewhere
along the trace. Moreover, both the MEG and HEG images should,
in principle, run through the position of the source. However,
in no cases, their intersection points with the readout trace 
coincide, due to systematic uncertainties. We chose the average 
of the two positions to obtain the coordinates of 
the source. The difference between the two provides an estimate 
of the systematic uncertainty associated with this technique, 
which is much larger than the statistical uncertainty. The 
results are summarized in Table 1. Note that for both GRS 1758-258
and Cyg X-1 the unusually small error bar in the Dec is due to 
the fact that the readout trace is nearly perpendicular to this 
direction; the error is, therefore, determined by the statistical 
uncertainty in fitting the cross linear profile of the trace.

The measured position of LMC X-1 is only about 0.7\arcsec\ away 
from the location of its optical counterpart (star 32; Cowley 
et al. 1978; Pakull 1978). The discrepancy is well within the 
uncertainty in the absolute aspect 
solutions\footnote{see http://cxc.harvard.edu/cal/ASPECT/celmon/}. 
Therefore, the result strongly supports the conclusions, based on 
dynamical mass determination, regarding the black hole candidacy 
of the source (Hutchings et al. 1987). For GRS 1758-258, the 
derived position is about 0.6\arcsec\ away from the position of 
its radio counterpart (whose is accurate to about 0.1\arcsec, 
Marti et al. 1998; also see Table~1). The association between the 
X-ray source and the core of the radio system (with prominent 
lobes) is, therefore, quite solid. Similarly, Cyg X-1 is 
determined to be about 0.5\arcsec\ away from its optical 
counterpart (HD 226868). On the other hand, the discrepancy 
is much larger for LMC X-3, which is about 2.7\arcsec\ away from 
its optical counterpart, although it could still be attributed 
to the uncertainty in the absolute aspect solutions. 

\subsection{Spectral Analysis}

A list of instrumenal features can be found in the {\it Proposers' 
Observatory Guide} (Table 9.4), which is available from the {\em 
Chandra X-ray Center}. All features in the LETG are also in the 
HETG (due to Au M edges) and the HETG has additional features at 
O K, Cr L and N K edges. Moreover, ACIS has an edge at 1.84 keV
(Si K) when the detector is a front-illuminated chip and 
the edge is very weak in a back-illuminated chip. 

To account for any remaining calibration uncertainties, we added 
1\% systematic uncertainty to the data for all sources. 

\subsubsection{LMC X-1}
We caught LMC X-1 in a very soft state. Fig.~2 shows the first-order 
HETG spectra of the source. The MEG and HEG spectra are shown separately 
for comparison. The spectral were rebinned so that there were at least 
15 counts in each energy bin. They were then jointly fitted with the 
MCD model (``diskbb'' in XSPEC) plus interstellar absorption (``wabs'' 
in XSPEC, using cross-sections from Morrison \& McCammon (1983)
and assuming solar abundances) and the quality of the fit is excellent. 
This seems to imply that the X-ray photons detected originate 
predominantly from the accretion disk; no Compton component is formally 
required to describe the data. The best-fit parameters are shown in 
Table~2. 

We tried but failed to detect any discrete features in the spectrum 
of LMC X-1. The absence of interstellar absoprtion edges is certainly 
due to low statistics of the data, since, after all, such 
features have been clearly seen in Cyg X-1 (Schulz et al. 2002). 
To be more quantitative, we derived upper limits on the 
optical depths of some of the most prominent edges (e.g., Ne K and 
Fe L3) detected in Cyg X-1. The procedure involved: replacing 
``wabs'' with ``varabs'' in the model; setting the abundance of
an element of interest to zero; and adding to the model an 
absorption edge (``edge'' in XSPEC) of the element (i.e., with
fixed energy). It should be noted that the ``varabs'' model uses
more recent cross sections from Balucinska-Church \& McCammon (1992).
We found that for the Ne K edge the 99\% upper limit is about 0.14, 
which is significantly lower than the 
expected value of 0.25 (derived by scaling the Cyg X-1 result by 
the best-fit column densities). On the other hand, the 99\% upper 
limit is about 0.90 for the Fe L3 edge, which is consistent with 
the scaled Cyg X-1 value (0.80). As for emission or absorption 
lines, the upper limits are not very constraining. For instance, 
we searched for Fe K$_{\alpha}$ emission similar to what was 
recently detected in 
Cyg X-1 (Miller et al. 2002). For such a line (which is centered at 
6.4 KeV and has a sigma width of 30 eV) the 99\% upper limit on 
the equivalent width is roughly 177 eV (compared to 16 eV as measured 
for Cyg X-1).

\subsubsection{LMC X-3}

The X-ray intensity of LMC X-3 is known to vary greatly. The variation 
was thought to be periodic and the period was determined to be $\sim$199 
or $\sim$99 days (Cowley et al. 1991). However, a simple periodic
modulation of the soft X-ray flux can be ruled out by the ASM/RXTE 
long-term monitoring of the source, although the flux seems to vary 
quasi-periodically on both timescales ($\sim$199 or $\sim$99 days).

LMC X-3 was in a very faint (and thus relatively hard) state when we 
observed it. Fig.~3 shows the first-order HETG spectra of the source, 
again with the MEG and HEG spectra shown separately for comparison. 
We proceeded to model the spectra in a manner that is similar to the
case of LMC X-1. Although the MCD model (plus interstellar 
absorption) produced a statistically acceptable fit to the data, 
the residuals show apparent features. An improved fit was achieved
by adding a Compton component (``comptt'' in XSPEC) to the model (see 
Fig.~3). The best-fit parameters are also summarized in Table~2.
The inferred HI column density is significantly lower than that toward 
LMC X-1, as is known from previous works. As in the case of LMC X-1, 
we failed to detect any discrete features in the spectrum. The absence
of interstellar absorption edges is certainly due to the combination 
of low column density along the line of sight and poor statistics of 
the data. The latter also prevents us from deriving meaningful 
constraints on emission or absorption lines. For example, the 99\% 
upper limit on the equivalent width of the 6.4 keV line is about 
643 eV!

\subsubsection{GRS 1758-258}

Like 1E1740.7-2942 (but unlike most BHCs), GRS 1758-258 was thought to 
always occupy the hard state (Tanaka \& Lewin 1995). Recently, however,
the source was seen to make a brief transition to the soft state as 
its X-ray flux decreased significantly (Smith et al. 2001). The 
phenomenon is rather unusual because the X-ray spectrum of a BHC 
usually hardens as its X-ray flux decreases. 

Fig.~4 shows the first-order HETG spectra of GRS 1758-258. The spectra 
are well described by the MCD model (plus interstellar absorption), 
which confirms the conclusions of Smith et al. (2001). The spectral 
fit results are shown in Table 2. The inferred hydrogen column density 
is in agreement with previous measurements (e.g., Lin et al. 2000 and
references therein). The derived temperature of the disk is in the 
typical range for BHCs in the soft state. In this case, two absorption
edges were positively detected at $1.308\pm 0.003$ keV and 
$1.845\pm 0.006$ keV and were thus attributed to Mg and Si K edges, 
respectively. The measured optical depths are $0.13 \pm 0.03$ and
$0.14 \pm 0.02$. The former (Mg) is significantly less than the scaled 
Cyg X-1 value but the latter (Si) is in agreement (see Schulz
et al. 2002 for values found in Cyg X-1). We found no emission or
absorption lines, confirming the results of Heindl \& Smith (2001) 
who had also
derived upper limits on the strength of a possible line over the
entire energy range. The spectrum of the source is so soft that there 
are very few counts at energies above 6 keV. Therefore, no meaningful 
constraints could be derived for the 6.4 keV emission.

\subsubsection{CYGNUS X-1}

The short-frame data is nearly pile-up free. Fig.~5 shows the 
first-order HETG spectra of Cyg X-1 which were constructed from such
data. Because of the very short effective exposure time (only about 
200 seconds) the quality of the data is insufficient to reveal the 
discrete features that are known to be present (Schulz et al. 2002). 

Cyg X-1 was also observed simultaenously with the Proportional Counter
Array (PCA) and High-Energy X-ray Timing Experiment (HEXTE) aboard
{\em RXTE}. The power density spectrum constructed from the PCA
data shows that the source was in the normal hard state, even though 
its soft flux seems a bit higher and spectrum softer (cf. Schulz et 
al. 2002). In terms of spectral coverages, the PCA significantly 
overlaps the HETG at low energies; the PCA and HEXTE extends the 
energy range up to about 200 keV. Since we were only interested in 
using the {\em RXTE} data to reliably determine the spectrum of 
non-disk emission, to minimize calibration uncertainties of the PCA 
at low energies we ignored data below 5 keV. Moreover, we only used 
data from the first xenon layer of each PCA detector unit. The PCA 
and HEXTE spectra are also shown in Fig.~5. For spectral modeling, 
we added 1\% systematic uncertainty to the {\em RXTE} data.

The {\em Chandra} and {\em RXTE} data were modeled jointly with 
the hydrogen column density fixed at $6.2\times 10^{21}\mbox{ }cm^{-2}$ 
(Schulz et al. 2002). We allowed the relative normalization between
the HETG and each of the PCA and HEXTE detector units to vary; the
results derived (e.g., fluxes and normalizations of individual 
spectral components) were based on the HETG normalization.
The usual combination of a MCD component and a 
Compton component failed to provide an adequate description of the 
data, even with the inclusion of a Gaussian function to account for 
residuals in the {\em RXTE} data between 5-8 keV. A much improved 
(but marginally acceptable) fit was achieved with a model consisting 
of two Compton components (plus the Gaussian function with its 
centroid energy fixed at 6.4 keV). There seem to be features in the
residuals mostly above 10 keV (see Fig.~5. The fit is nevertheless 
good enough for decoupling the soft and hard components, so we 
refrained ourselves from trying more sophisticated models invoking 
disk reflection. The best-fit parameters are summarized in Table~3. 
The Gaussian feature is broad 
($\sigma \sim 1$ keV) and its equivalent width is about
150 eV. It is intriguing that these properties are similar to
those of the disk line reported by Miller et al. (2002), although
complicated calibration issues regarding the PCA often make it 
difficult to assess the reality of such features. We will not discuss 
it any further. 

\section{Summary and Discussion}

Unlike Cyg X-1 (Schulz et al. 2002; Marshall et al. 2001; Miller 
et al. 2002) or 1E1740.7-2942 (Cui et al. 2001), we found no 
evidence for the presence of any emission or absorption lines in the 
spectra of LMC X-1, LMC X-3, and GRS 1758-258. This is at least
partly due to the lack of statistics, as evidenced by the absence
of interstellar absorption edges. In the case of GRS 1758-258, we 
did positively detect Mg K and Si K edges, which are most likely 
of interstellar origin. 

For BHCs, X-ray illumination of the accretion disk is thought to
be very important and is thus necessarily taken into account in 
the models. The re-processing of incident X-rays by the disk is 
often used to explain some of the continuum features. This 
process is expected to be accompanied by the production of 
fluorescent lines. On the other hand, depending on the ionization
state of the disk, the lines can be destroyed by the Auger process
(Ross \& Fabian 1993). The lack of line emission can, therefore,
shed light on these physical processes. Unfortunately, in our case,
the quality of data is not good enough to constrain the models. 

We have shown that for most sources in our sample the observed X-ray 
continuum can be well described by a two-component model including
a MCD component and a Compton component. Such a model has been 
successfully applied to BHCs in general previously. The low-energy 
sensitivity of the HETG has allowed us to reliably model the 
emission from the hot inner region of the accretion disk (i.e., the
MCD component). As in previous studies, however, the coupling 
between the two components still makes it challenging to completely
separate them. Luckily, in two cases, LMC X-1 and GRS 1758-258, the 
data requires only the MCD component, indicating the X-ray photons
from these sources originate entirely from the disk. Here, the 
derived disk parameters should, in principle, be quite reliable.

There are three important issues that make it uncertain to extract
the physical properties of the accretion disk from the measured
quantities (e.g., Zhang 1999; Cui 2001). First of all, it is still 
not clear what is the origin of seed
photons for the inverse Comptonization process. If the accretion 
disk provide all the seed photons, as often assumed, these photons
will not appear in the observed spectrum of the disk (i.e., the MCD
component). Instead, they appear in the Compton component. In this
scenario, therefore, the photons detected are originally all
from the accretion disk. To compute intrinsic luminosity of the 
disk we must then add photons in the Compton component to the MCD
component. It is the intrinsic disk luminosity that should be used
to derive such important properties as the radius of the inner edge
of the disk. In practice, however, the disk radius is often derived
directly from the best-fit normalization of the MCD component, which 
is clearly erroneous, especially when the Compton component is seen 
to be strong. This erroneous procedure is likely to be responsible
for the claimed correlation between the inner disk radius and the
spectral hardness of the source: the stronger the Compton component
the smaller the radius, since this is exactly what would be expected 
if the disk is the primary source of seed photons for Comptonization. 
Unfortunately, the derivation of the intrinsic luminosity of the 
disk (by taking into account the loss of disk photons to the 
Comptonization process) depends on many assumption, such as the 
exact origin of seed photons and the spatial distribution of 
Comptonizing electrons with respect to the disk, and is thus very 
model dependent. Robust results can be obtained only for cases in 
which the Compton component is either absent or weak enough to 
produce negligible effects.

Secondly, as discussed in \S~1, the local spectrum of the inner 
portion 
of the disk is not a blackbody, because the opacity is predominantly 
determined by electron (Compton) scattering as opposed to free-free 
emission (e.g., Ebisuzaki, Hanawa, \& Sugimoto 1984). Strictly speaking, 
therefore, the spectral shape of the disk emission is that of a 
(saturated) Comptonized spectrum, with the spectrum of seed photons 
being of MCD shape. The ratio of the temperature of electrons in the 
disk to that of seed photons gives the so-called ``color correction
factor'' (or more precisely, spectral hardening factor). This factor 
is critical to deriving the radius of
the inner edge of the disk (see discussion in \S~1). A possible way
to determine the hardening factor observationally is, therefore, to
model the disk spectrum with a Compton component. We investigated
this possibility by replacing the MCD component (see Table 2) in
the model with ``comptt''. The new model fit the data equally well, 
if not better, for all cases. The results are summarized in Table 3. 
As a sanity check, we expect that the temperature ($T_e$) of 
Comptonizing electrons in the disk should be close to the effective
temperature of the disk, when the optical depth is large. This is
indeed the case (comparing results in Table 3 and Table 2), although
the measured values of the optical depth are much smaller than that 
expected from the standard $\alpha$ disk model (Shakura \& Sunyaev 
1973). It should be noted that ``comptt''
assumes a Wien (as opposed to MCD) spectrum for seed photons 
(Titarchuk 1994). Empirically, we found that in order to fit the 
peak of a MCD spectrum with a Wien function the temperature of
the inner disk ($T_{in}$ in MCD) must be equal to 2.7 times that
of the Wien distribution ($T_0$). Therefore, the spectral hardening
factor is simply given by $f = T_e/(2.7T_0)$. With large uncertainties
(mostly due to poor constraints on the seed photon temperature), the
values of $f$ were found to be $2.6$, $\gtrsim 0.3$, $1.6$, and $4.7$, 
respectively, for the sources in Table~3. While some of these values 
seem reasonable (compared to the results of Merloni et al. 2000), 
others are too high (but keep in mind the large uncertainties).  
Data of much improved statistics, especially at low energies, is 
required to constrain the temperature of seed photons.

Finally, the situation can be further complicated by the presence 
of a ``warm layer'' just above the disk (Zhang et al. 2000). Such 
a layer can be due to the heating of the disk by an illuminating
hard X-ray source (e.g., Nayakshin \& Melia 1997; Mistra et al. 
1998), although observation evidence suggests that the layer 
exists even in the absence of any non-disk emission (Zhang et al. 
2000). The presence of a low-density warm layer is in fact supported 
by our results (i.e., relatively small optical depth of the soft 
Compton component for all sources; see Table 3). To account for the 
warm layer, yet another Compton component needs to be added to the 
model, which is not warranted by the quality of our data. This 
can perhaps explain somewhat erratic values of the spectral 
hardening factor.

To conclude, we emphasize that although modeling the disk
continuum is a viable approach to deriving physical parameters of 
the disk the derivation is complicated by the issues discussed. 
Progress can be made by improving the quality of the data, 
especially at the lowest energies. Equally important is simultaneous 
broadband coverages that make it possible to reliably separate
out the disk component from other components. The value of broad
spectral coverage was demonstrated by the modeling of the Cyg X-1 
spectrum: the traditional MCD model was shown to fail in this 
case. 

\acknowledgments

We wish to thank Mark Ertmer and Joe Germano for their assistance with 
data analysis and Herman Marshall for providing information on HETG
calibrations. We also acknowledge useful discussions with Yangsen
Yao and Xiaoling Zhang on the subject of spectral color correction. 
This work was supported in part by NASA through Contract 
NAS8-38252 (with subcontracts to MIT and Purdue) and the LTSA grant 
NAG5-9998.

\clearpage
\begin{deluxetable}{lll}
\tablecolumns{3}
\tablewidth{0pc}
\tablecaption{{\em Chandra} Positions of the Sources\tablenotemark{1}}
\tablehead{
\colhead{Source} & \colhead{RA (J2000)} & \colhead{Dec (J2000)}
}
\startdata
LMC X-1  &  $05^h\mbox{ }39^m\mbox{ }38^s.85 \pm 0.02$ &
$-69$\arcdeg\ $44$\arcmin\ $35$\arcsec$.71 \pm 0.02$ \\
star 32\tablenotemark{2} & $05^h\mbox{ }39^m\mbox{ }38^s.72$ &
 $-69$\arcdeg\ $44$\arcmin\ $35$\arcsec$.6$ \\ \hline
LMC X-3  &  $05^h\mbox{ }38^m\mbox{ }56^s.63 \pm 0.05$ &
$-64$\arcdeg\ $05$\arcmin\ $03$\arcsec$.29 \pm 0.08$ \\
WP8\tablenotemark{3} &$05^h\mbox{ }38^m\mbox{ }56^s.4$ & 
$-64$\arcdeg\ $05$\arcmin\ $01$\arcsec \\ \hline
GRS 1758-258 & $18^h\mbox{ }01^m\mbox{ }12^s.43 \pm 0.02$ &
$-25$\arcdeg\ $44$\arcmin\ $35$\arcsec$.524 \pm 0.005$ \\
VLA C\tablenotemark{4} & $18^h\mbox{ }01^m\mbox{ }12^s.395$ & 
$-25$\arcdeg\ $44$\arcmin\ $35$\arcsec$.90$ \\ \hline
Cyg X-1 & $19^h\mbox{ }58^m\mbox{ }21^s.656 \pm 0.006$ &
$+35$\arcdeg\ $12$\arcmin\ $06$\arcsec$.370 \pm 0.005$ \\ 
HD226868\tablenotemark{5} &$19^h\mbox{ }58^m\mbox{ }21^s.6756$ & 
$+35$\arcdeg\ $12$\arcmin\ $05$\arcsec$.775$ \\
\tablenotetext{1}{The table also lists the optical or radio counterpart 
of each source.}
\tablenotetext{2}{Cowley et al. 1978; Pakull 1978}
\tablenotetext{3}{Warren \& Penfold 1975; Cowley et al. 1978}
\tablenotetext{4}{Rodr\'\i guez, Mirabel, \& Mart\'\i\ 1992}
\tablenotetext{5}{The coordinates shown were taken from SIMBAD (based
on the Hipparcos catalogue).}
\enddata
\end{deluxetable}

\begin{deluxetable}{lccccccccc}
\rotate
\scriptsize
\tablecolumns{10}
\tablewidth{0pc}
\tablecaption{Best-Fit Parameters: diskbb+compTT\tablenotemark{1} }
\tablehead{
 & & \multicolumn{2}{c}{Multi-Color Disk\tablenotemark{2}} & \multicolumn{4}{c}{Comptonized Hard Component\tablenotemark{3}} \\
\cline{3-4} \cline{5-8} \\
\colhead{Source} & \colhead{$N_H$} & \colhead{$kT_{dbb}$}& \colhead{$N_{dbb}$} & \colhead{$kT_0$} & \colhead{$kT_e$} & \colhead{$\tau$} & \colhead{$N_{comp}$} & \colhead{$\chi^2_{\nu}/dof$} & \colhead{Flux\tablenotemark{4}}\\
 & $10^{22}\mbox{ }cm^{-2}$ & keV & & keV & keV & & & & }
\startdata
LMC X-1 & $0.46_{-0.02}^{+0.02}$ & $1.10_{-0.01}^{+0.01}$ & $15.5_{-0.7}^{+0.7}$ & \nodata & \nodata & \nodata & \nodata & $0.7/4161$ & $3.13$ \\
LMC X-3 & $< 0.06$ & $0.77_{-0.04}^{+0.07}$ & $16_{-5}^{+4}$ & $0.12_{-0.03}^{+0.02}$ & $66_{-59}^{+391}$ & $0.9_{-0.7}^{+1.9}$ & $0.001_{-0.001}^{+0.005}$& $0.5/2250$ & $1.70$ \\
GRS 1758-258 & $1.83_{-0.03}^{+0.03}$ & $0.50_{-0.01}^{+0.01}$ & $539_{-45}^{+45}$ &\nodata& \nodata & \nodata & \nodata & $0.7/2966$ & $1.07$ \\
\tablenotetext{1}{The uncertainties shown represent 90\% confidence intervals.}
\tablenotetext{2}{$T_{dbb}$ and $N_{dbb}$ are the temperature and 
normalization, respectively.}
\tablenotetext{3}{$T_0$ is the seed photon temperature; $T_e$ is the 
electron temperature; $\tau$ is optical depth; and $N_{comp}$ is the
normalization.}
\tablenotetext{4}{Observed 0.5--10 keV keV flux in units of $10^{-10}\mbox{ }erg\mbox{ }cm^{-2}\mbox{ }s^{-1}$.} 
\enddata
\end{deluxetable}

\begin{deluxetable}{lcccccccccc}
\rotate
\scriptsize
\tablecolumns{11}
\tablewidth{0pc}
\tablecaption{Best-Fit Parameters: compTT+compTT\tablenotemark{1} }
\tablehead{
 & & \multicolumn{4}{c}{Comptoninzed Disk Component} & \multicolumn{4}{c}{Comptonized Hard Component} \\
\cline{3-6} \cline{7-10} \\
\colhead{Source} & \colhead{$N_H$} & \colhead{$kT_0^d$}& \colhead{$kT_e^d$} & \colhead{$\tau^d$} & \colhead{$N_{comp}^d$} & \colhead{$kT_0$} & \colhead{$kT_e$} & \colhead{$\tau$} & \colhead{$N_{comp}$} & \colhead{$\chi^2_{\nu}/dof$} \\
 & $10^{22}\mbox{ }cm^{-2}$ & keV & keV & & & keV & keV & & & }
\startdata
LMC X-1 & $0.59_{-0.03}^{+0.03}$ & $0.13_{-0.13}^{+0.04}$ & $0.92_{-0.03}^{+0.03}$ & $13_{-1}^{+1}$ & $0.36_{-0.02}^{+0.02}$ & \nodata & \nodata & \nodata & \nodata & $0.7/4159$ \\
LMC X-3 & $0.05_{-0.05}^{+0.06}$ & $< 0.75$ & $0.536_{-0.009}^{+0.009}$ & $70_{-7}^{+8}$ & $0.059_{-0.003}^{+0.003}$ & $0.122_{-0.008}^{+0.008}$ & $86_{-4}^{+5}$ & $0.43_{-0.02}^{+0.03}$ & $1.78_{-0.05}^{+0.06} \times 10^{-3}$& $0.5/2244$ \\
GRS 1758-258 & $2.07_{-0.14}^{+0.14}$ & $0.11 _{-0.11}^{+0.08}$ & $0.47_{-0.02}^{+0.02}$ & $12.4 _{-1.6}^{+3.2}$ & $3.8_{-1.2}^{+1.2}$ &\nodata& \nodata & \nodata & \nodata & $0.7/2964$ \\
Cyg X-1\tablenotemark{2} & $0.62$  & $0.10_{-0.02}^{+0.02}$ & $1.24_{-0.06}^{+0.06}$ & $6.1_{-0.3}^{+0.4}$ & $27_{-7}^{+12}$ & $0.97_{-0.04}^{+0.05}$ & $34_{-2}^{+1}$ & $3.9_{-0.2}^{+0.1}$ & $4.4_{-0.3}^{+0.2} \times 10^{-2}$ & $1.5/1789$ \\
\tablenotetext{1}{The same as Table 2, unless otherwise noted.}
\tablenotetext{2}{In this case, the column density is fixed and the 
fit also requires a Gaussian function with its centroid fixed at 
6.4 keV. The Gaussian function has a width of 0.94 keV and an 
equivalent width of 150 eV. }
\enddata
\end{deluxetable}

\clearpage

\begin{figure}                                                           
\psfig{figure=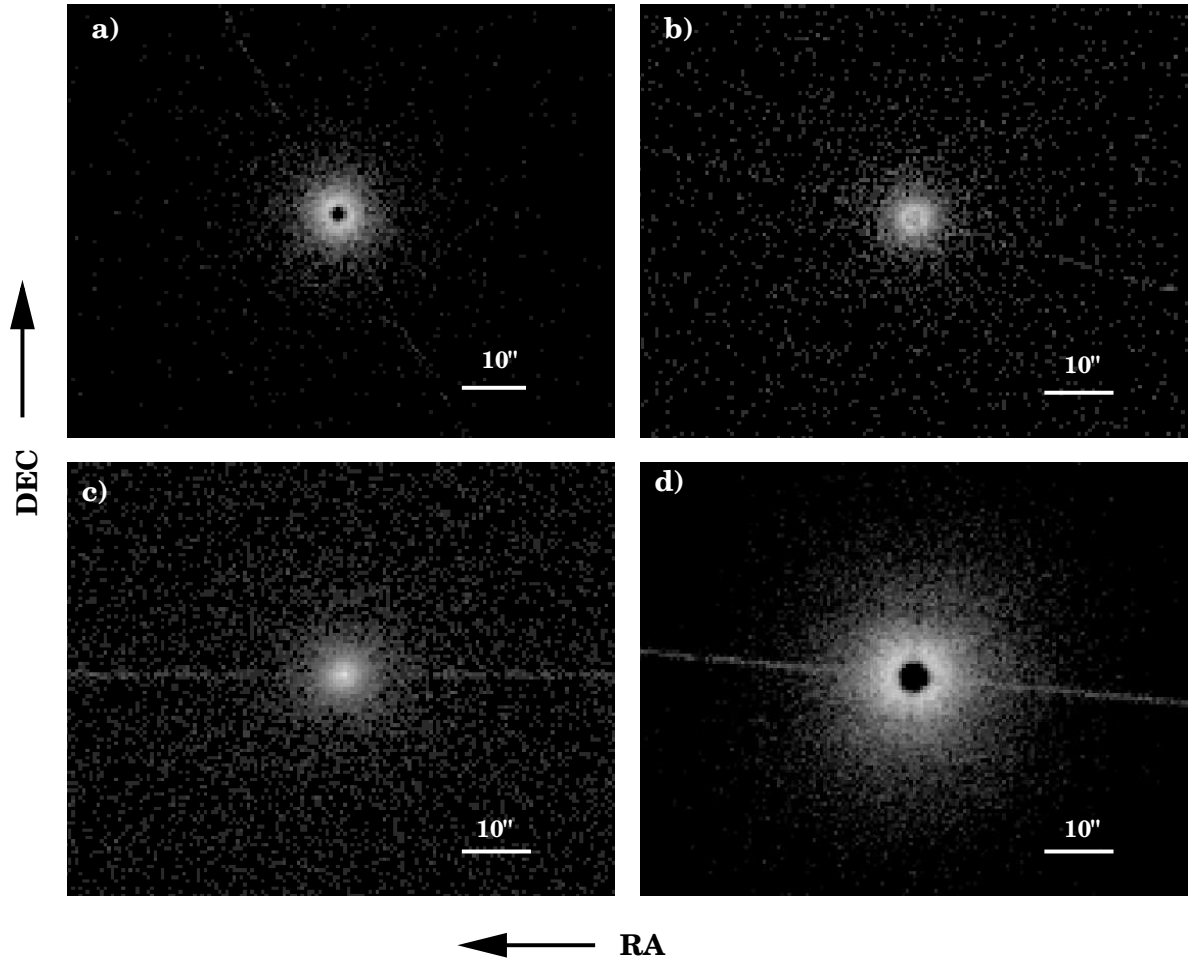,width=5in,angle=-90}
\caption{Unsmoothed zeroth-order images of the sources and 
their surroundings: (a) LMC X-1, (b) LMC X-3, (c) GRS 1758-258, and 
(d) CYG X-1. The images are roughly 1\arcmin$\times$1.26\arcmin\ (with
0.5\arcsec$\times$0.5\arcsec\ pixels). The intensity of each image is scaled 
logarithmically. The presence of ``scattering halos'' (see text) is 
apparent. Note the effects due to severe photon pileup in three of the 
cases, in which a ``valley'' appears at the position of the source. Also,
the CCD readout traces are not removed from the images. }
\end{figure}

\begin{figure}                                                           
\psfig{figure=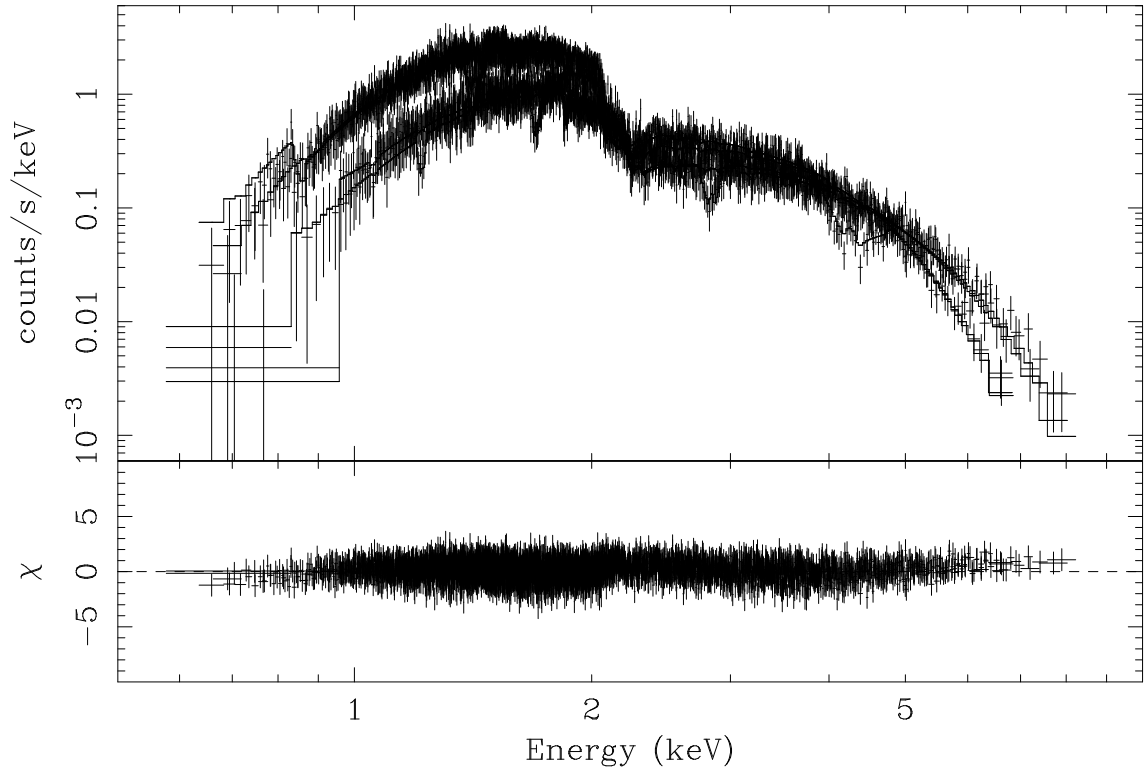,width=4in,angle=-90}
\caption{First-order HETG count spectra of LMC X-1. The plus and minus 
orders of the MEG and HEG are shown separately. The folded model curves 
are shown in solid histograms. }
\end{figure}

\begin{figure}                                                           
\psfig{figure=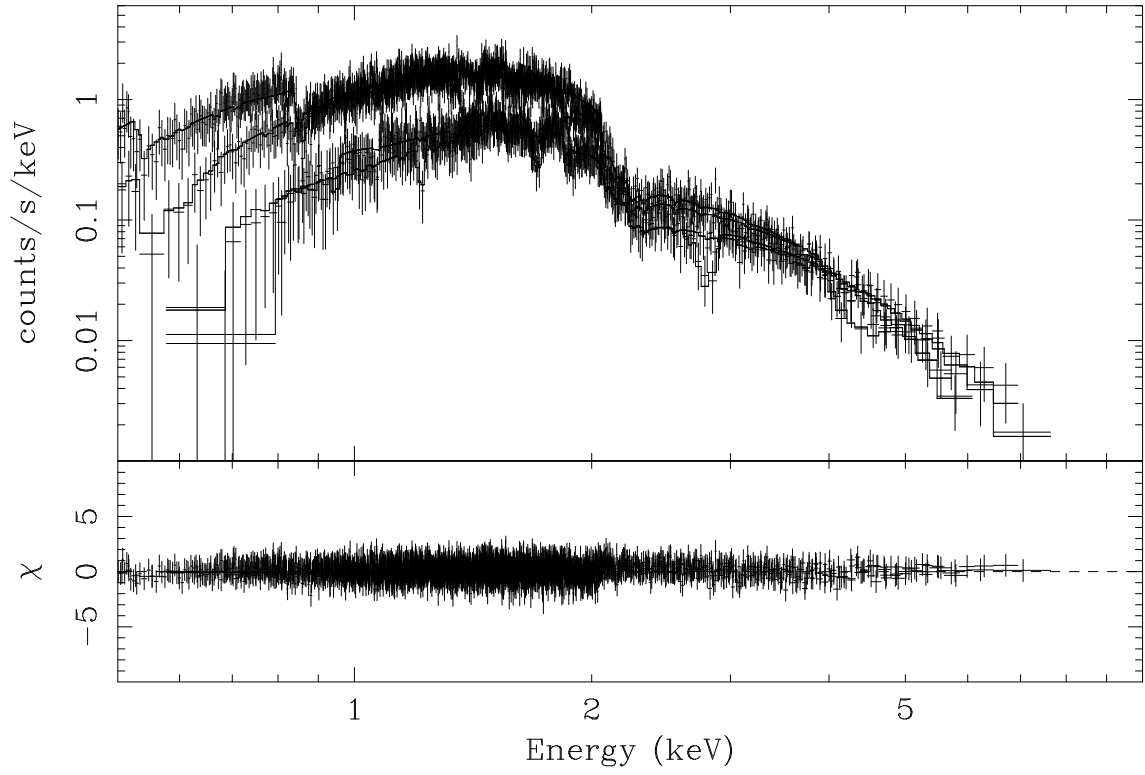,width=4in,angle=-90}
\caption{Same as Fig. 2 but for LMC X-3.}
\end{figure}

\begin{figure}                                                           
\psfig{figure=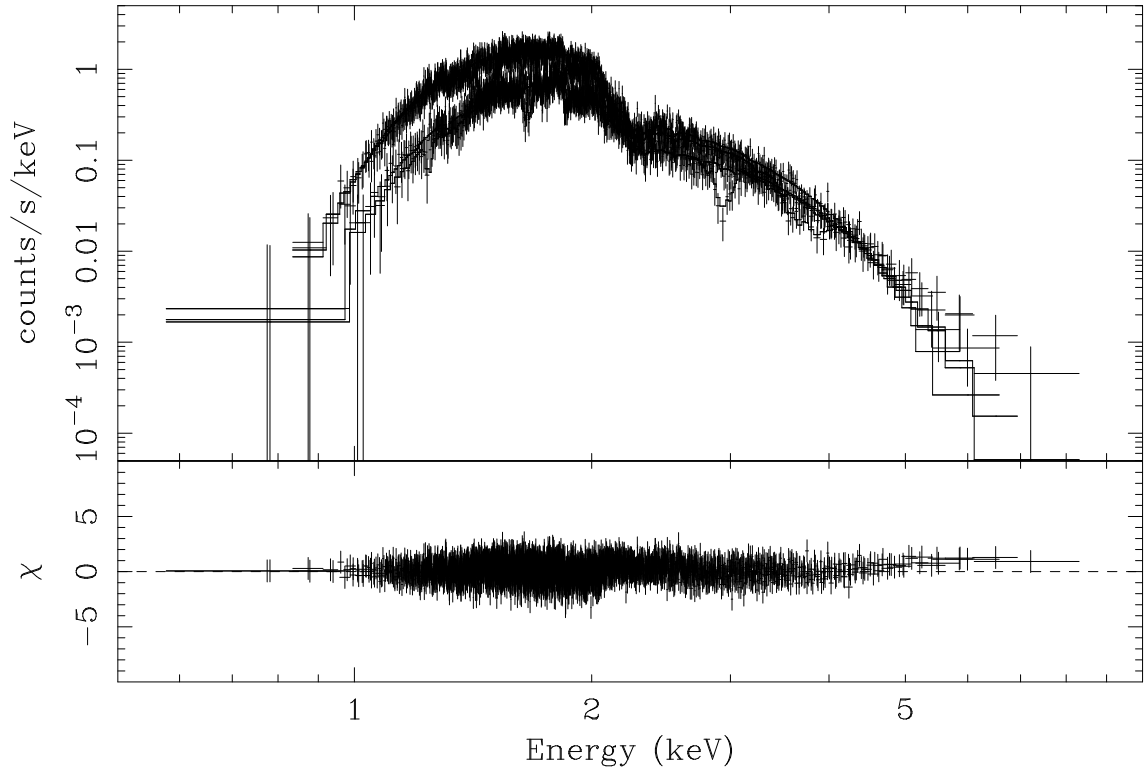,width=4in,angle=-90}
\caption{Same as Fig. 2 but for GRS 1758-258.}
\end{figure}

\begin{figure}                                                          
\psfig{figure=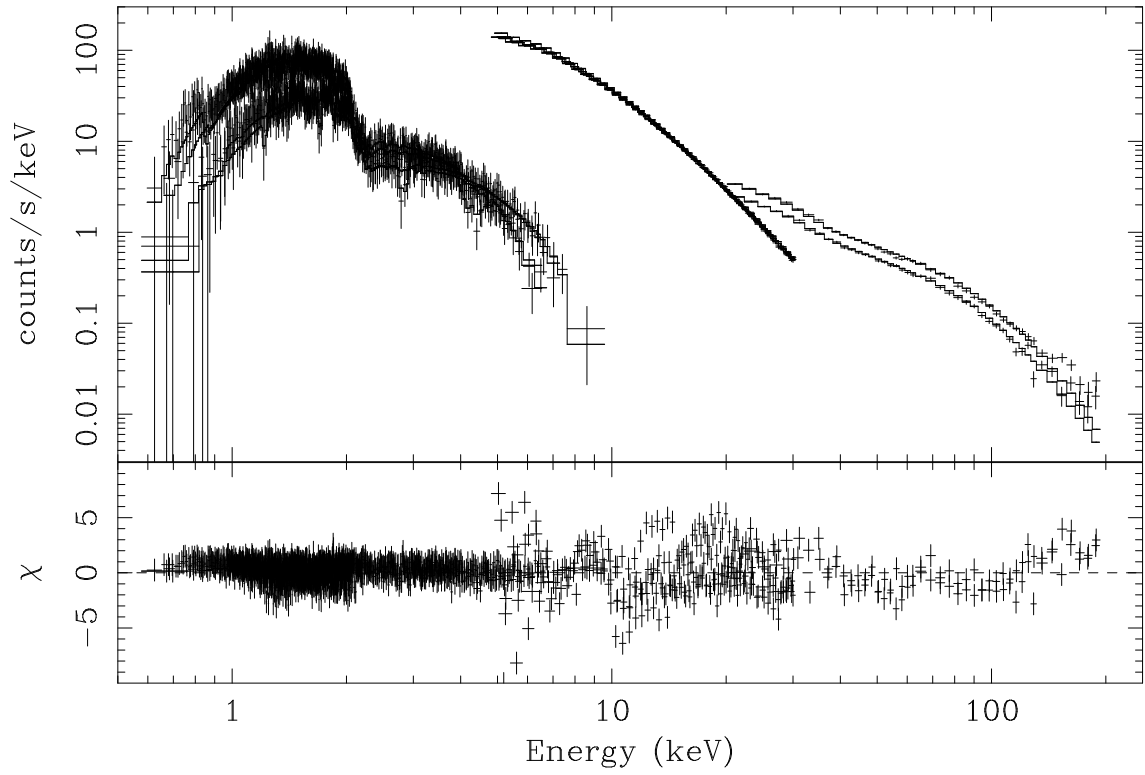,width=4in,angle=-90}
\caption{Same as Fig. 2 but for CYG X-1. Also shown are the PCA and
HEXTE spectra of the source from simultaneous observations with
{\em RXTE}. }
\end{figure}

\end{document}